\documentclass[oneside,a4paper,11pt,shownumbers]{article}
\usepackage{latexsym}
\usepackage{euscript}
\usepackage{epsfig,amsmath,amssymb}
\usepackage[dvips]{color}

\newcommand{\lb}{\label}

\def\be{\begin{equation}}
\def\ee{\end{equation}}
\def\bea{\begin{eqnarray}}
\def\eea{\end{eqnarray}}

\long\def\symbolfootnote[#1]#2{\begingroup%
\def\thefootnote{\fnsymbol{footnote}}\footnote[#1]{#2}\endgroup}

 \large\normalsize

\begin{document}

\title{\LARGE \bf Bouncing Cosmological Isotropic Solutions in \\Scalar-Tensor Gravity}
\author{ D.~Polarski$^1$\thanks{david.polarski@umontpellier.fr},
~A.~A.~Starobinsky$^{2,3}$\thanks{alstar@landau.ac.ru},
~Y.~Verbin$^4$\thanks{verbin@openu.ac.il} }
 \date{\today}
 \maketitle
 \vskip -1.0cm
\centerline{$^1$\em Universit\'e de Montpellier \& CNRS, Laboratoire Charles Coulomb,}
\centerline{\em UMR 5221, F-34095 Montpellier, France}
\vskip 10pt
\centerline{$^2$\em L. D. Landau Institute for Theoretical Physics RAS,}
\centerline{\em Moscow 119334, Russian Federation}
\vskip 10pt
\centerline{$^3$\em Kazan Federal University, Kazan 420008, Republic of Tatarstan, Russian Federation}
\vskip 10pt
\centerline{$^4$\em Astrophysics Research Center \& Department of Natural Sciences,}
\centerline{\em  The Open University of Israel, Raanana 4353701, Israel}

\maketitle
\thispagestyle{empty}

\begin{abstract}
Bouncing non-singular isotropic cosmological solutions are investigated in a simple model of scalar-tensor gravity.
%
New families of such solutions are found and their properties are presented and analyzed
using an effective potential as the main tool.
Bouncing solutions are shown to exist for a Higgs-like self-interaction
potential which is bounded from below,
in contrast to previous solutions that appeared in the literature based on potentials which
were unbounded from below.
In the simplest version of a scalar field with the quartic potential and conformal
coupling to gravity, bouncing spatially flat solutions either have the Hubble function diverging in the past before
the bounce, but with a well-behaved future, or are
globally regular but unstable with respect to anisotropic or inhomogeneous
perturbations at some finite values of the scalar field and curvature. 
Regular solutions can only exist in the part of the parameter space where the maximum of
the effective potential is larger than the first zero of the potential, and gravity becomes
repulsive at the bounce.
%
%
%

\end{abstract}

\maketitle
\medskip \medskip

\section{Introduction}
\setcounter{equation}{0}
Inflation as the stage of the early Universe evolution preceding the hot Big Bang (the radiation dominated stage) is widely considered to be the most elegant and promising idea to provide definite initial conditions for the 
 evolution of the Universe up to the present time. And indeed, a narrow subclass of inflationary models, the simplest of them depending on one free parameter to be fixed by observations only, quantitatively predicted the form of the observed spectrum of primordial scalar (matter density) inhomogeneities in the Universe at large scales, while many other models have been already excluded mainly by the measured value of the slope of the power spectrum of primordial curvature perturbations $n_s(k)-1$~\cite{Planck:2018vyg} and by the upper value of the tensor-to-scalar ratio $r(k)$ (see~\cite{BICEP:2021xfz} for the most recent strong upper limit). However, the duration of the inflationary stage inside our past light cone is finite in all these viable inflationary models. Thus, even if we accept the existence of the inflationary stage in the past of our Universe, it is natural to think what was before it.
One possibility is that inflation was preceded by a generic inhomogeneous and anisotropic stage with curvature much exceeded that during inflation (even up to singular values). Indeed in this case it is straightforward to show that inflation is an intermediate attractor and it can be realized for a set of initial pre-inflationary conditions with a nonzero measure\footnote{In this case the fact that inflationary models do not 'solve' the singularity problem (that they did not intend to), so they permit more general solutions with much higher curvature, actually serves as an advantage, not a drawback.}. For the idealized cases of inflation in General Relativity (GR) with a cosmological constant and power-law inflation where inflation is the generic late-time attractor, this was proved in \cite{Starobinsky:1982mr} and \cite{Muller:1989rp} respectively, see also the recent papers \cite{Muller:2017nxg,Mishra:2019ymr} for anisotropic cosmological solutions in the more realistic $R+R^2$ and Higgs inflationary models.

However, another possibility considered in this paper is presented by bouncing cosmologies which have a regular minimum of the scale factor at some finite value of curvature much less than the Planck one. It is easy to obtain such solutions even in the Friedmann-Lema\^ itre-Robertson-Walker (FLRW) case in the presence of spatial curvature. For example, in \cite{S78} the bouncing FLRW model with a massive scalar field minimally coupled to gravity and positive spatial curvature was constructed which, in fact, had both a contracting quasi-de Sitter stage before the bounce and an expanding slow-roll quasi-de Sitter stage after it. However, though this model permits solutions with an arbitrary large number of bounces, completely regular solutions with an infinite number of bounces exist in it, too, but they are not generic~\cite{Page:1984qt,Kamenshchik:1997szb,Kamenshchik:1998xx,Kamenshchik:1998ix}. So, this bouncing model does not solve the singularity problem. Many other bouncing models in theories of modified gravity including scalar-tensor gravity were proposed later, see in particular
\cite{Mukhanov-Brandenberger1991,Novello-Bergliaffa2008,BrandenbergerPeter2016,Barrow:2017yqt} and the recent papers~\cite{Bramberger:2019zez,Ganguly:2019llh,Gungor:2020fce}.
Such theories are an ideal framework for the realization of nonsingular
bouncing cosmologies, since they allow for the violation of the null energy
condition that is necessary for the existence of a bounce in the absence of spatial curvature \cite{Boisseau:2000pr,NojiriOdintsov2006,CapozzielloDeLaurentis2011,NojiriEtAl2017}. In particular,
one can obtain bouncing solutions in pre-big-bang
\cite{Veneziano1991} and ekpyrotic \cite{Khoury+3-2001,Khoury+4-2001} models
(though with some additional assumptions on matching of solutions before and after the bounce).

It is just the case of FLRW bouncing models with zero spatial curvature on which we will concentrate in this paper. In a series of three recent papers \cite{BoisseauEtAl1,BoisseauEtAl2,BoisseauEtAl3} by
Boisseau, Giacomini, Polarski and Starobinsky (BGPS), bouncing cosmological solutions were
found in a simple field theoretical model:
%
%
BGPS took the simple potential $U(\Phi)=U_0 + \lambda \Phi^4 /4$ and found
that in order to obtain bouncing cosmological solutions, the cosmological term $(U_0)$ should
be positive and the self-coupling ($ \lambda$) should be negative, thus rendering the potential
unbounded from below. The case of $\xi = 1/6$ which corresponds to conformally coupled scalar,
results a considerable simplifications in the field equations and allows for explicit
(analytic) solutions \cite{BoisseauEtAl1}.
The BGPS bouncing solutions start at a point in the past where the scalar field has a maximal
value, from which it decreases while the scale factor $a(t)$ decreases as well and passes
through a minimum. The scalar field keeps decreasing and vanishes asymptotically, while the
scale factor starts increasing exponentially with an asymptotically constant Hubble function
$H=\dot{a}/a$. As usual the bounce point is defined in terms of the Hubble function as the
time where $H=0$ and $\ddot{a}>0$.
Some further progress in this direction was made more recently
\cite{KamenshchikEtAl2015,PozdeevaEtAl2016} still using unbounded from below potentials.
%
%
%
Finally, we must mention the problem of the negative effective gravitational constant
\cite{Starobinsky1981} (see also \cite{ABGS03}) which is obvious if we notice that the coefficient of the
Einstein-Hilbert term in the action (\ref{totalAction}) is the field dependent factor
$1/\kappa^2 - \xi \Phi^2$.
This factor vanishes for a definite field value, but as we will
see, homogeneous and isotropic cosmological solutions cross this field value without any
noticeable difficulty. However, small deviations from isotropy
\cite{Starobinsky1981,Futamase-Maeda1987,FutamaseEtAl1989} or from homogeneity
\cite{CaputaEtAl2013} diverge as the scalar field tends to $\Phi^2 \rightarrow 1/\xi\kappa^2$.
For the sake of completeness of the discussion, we choose first to present the homogeneous
and isotropic cosmological solutions in the entire domain where they exist without imposing
the further restriction on inhomogeneous and anisotropic instabilities.
The paper is organized in the following way. In Section 2 we present the scalar-tensor model
 that we study in this paper written in the Jordan frame in a general context which will allow further extensions. In Section 3 we specialize to the case of conformal coupling which results in considerable simplifications of the equations. We obtain the effective
potential of the scalar field and the conditions on the (quartic) potential
parameters $m^2,\lambda$ are derived in order to obtain bouncing solutions. In Section
4 the relevant regions in this parameter space are found, in particular that region
where a negative effective gravitational constant around the bounce can be avoided, and
we discuss the properties of bounce solutions found there. The region where all quantities
remain regular at all times, corresponds to that region where the effective gravitational
constant necessarily becomes negative. In section 5 we analyze the problem in the Einstein
frame and we recover the region where bounce solutions avoiding a negative effective
gravitational constant around the bounce are possible. Our results are summarized and
discussed briefly in our Conclusion.
\section{Scalar-Tensor Gravity: Action and Field Equations}
\setcounter{equation}{0}

We consider a scalar-tensor gravity model whose action is given by
\be
S = \int d^4 x \sqrt{- g} \left (\frac{R}{2 \kappa^2} +
             \frac{1}{2}\partial_{\mu}\Phi \partial^{\mu} \Phi
              - U(\Phi) -\frac{\xi}{2} R \Phi^2 \right )~, \label{totalAction}
\ee
where  $R$ is the Ricci scalar\footnote{We will use the $(+,-,-,-)$ signature, while our definition
of $R$ has a minus sign compared to Landau-Lifshitz.} and $\Phi$ is a real scalar field
non-minimally coupled to gravity through the coupling constant $\xi$.
Here $\kappa^2 = 8 \pi G$, with $G$ the usual gravitational (Newton's) constant.
The action \eqref{totalAction} is actually a special case of the general family which can be written in the notation of \cite{BoisseauEtAl1} as
\be
S = \frac{1}{2}\int d^4 x \sqrt{- g} ~ \left( F(\Phi)R
        + Z(\Phi)\partial_{\mu}\Phi \partial^{\mu} \Phi - 2U(\Phi) \right)~,\lb{actionJ}
\ee
where we choose here
\be
F(\Phi) = \kappa^{-2}-\xi \Phi^2 ~,\label{F}
\ee
and
\be
Z(\Phi) = 1~. \lb{Z=1}
\ee
It is obvious that by a field redefinition the function $Z(\Phi)$ can be reduced to be $\pm1$. We note further that $Z(\Phi) = -1$
is another possibility which is perfectly acceptable provided the corresponding Brans-Dicke
parameter (function) $\omega_{\rm BD}$ satisfies $-\frac32 < \omega_{\rm BD} < 0$.
We will consider however models with $Z(\Phi) = 1$.

The corresponding field equations of the general theory \eqref{actionJ} with $Z(\Phi) = 1$ read
\be
\nabla_{\mu} \nabla^{\mu} \Phi + U'(\Phi) -\frac{R}{2}  F'(\Phi ) =0 \;\;\ , \;\;\;\
F(\Phi ) G_{\mu\nu} +  T^{(min)}_{\mu\nu} + \nabla_{\mu}\nabla_{\nu} F(\Phi) -
g_{\mu\nu}\nabla_{\rho} \nabla^{\rho} F(\Phi)=0
\label{FEqScalar+Grav_Gen}
\ee
where $T^{(min)}_{\mu\nu}$ is the energy-momentum tensor of the minimally coupled scalar field:
\be
T^{(min)}_{\mu\nu}=\partial_{\mu}\Phi \partial_{\nu} \Phi -   \left ( \frac{1}{2}\partial_{\rho}\Phi \partial^{\rho} \Phi - U(\Phi) \right)g_{\mu\nu}
~.\label{Tmin}
\ee

In this work we would like to concentrate on the scalar field dynamics in the simplest case \eqref{F}.
Writing the field equations for the scalar field, we find
\be
\nabla_{\mu} \nabla^{\mu} \Phi + U'(\Phi) +\xi R \Phi =0~,  \lb{KGa}
\ee
while we obtain for the gravitational field
\be
(1-\xi \kappa^2 \Phi^2)G_{\mu\nu} + \kappa^2 T^{(min)}_{\mu\nu} +
     2\xi \kappa^2 \left ( \partial_{\rho}\Phi \partial^{\rho} \Phi +
        \Phi \nabla_{\rho} \nabla^{\rho} \Phi  \right)g_{\mu\nu} -
     2\xi \kappa^2 \left ( \partial_{\mu}\Phi \partial_{\nu} \Phi +
        \Phi \nabla_{\mu} \nabla_{\nu} \Phi  \right) = 0~,
\label{FEqGrav}
\ee
where $T^{(min)}_{\mu\nu}$ is given by \eqref{Tmin}.
%
By tracing Eq. \eqref{FEqGrav} we obtain an expression for $R$ in terms of the scalar field
\be
R=\kappa^2 \frac{ 4 U(\Phi) -6 \xi \Phi U'(\Phi) - (1-6 \xi)(\partial\Phi)^2}{1-\xi \kappa^2  (1-6 \xi)\Phi^2}
\label{RicciPhiGenXi}
\ee
where we use the notation $(\partial\Phi)^2 = \partial_{\rho}\Phi \partial^{\rho} \Phi$ . So the equation of the scalar field \eqref{KGa} may be simplified to be written in terms of an effective potential $U_{\rm eff}(\Phi)$ as
\be
\nabla_{\mu} \nabla^{\mu} \Phi - \frac{ \xi \kappa^2 (1-6 \xi)\Phi}{1-\xi \kappa^2  (1-6 \xi)\Phi^2} (\partial\Phi)^2 + U_{eff}'(\Phi) =0
\label{FEqScalarEffPotGenXi}
\ee
where the effective potential is defined in terms of its derivative\footnote{This effective
potential is different from the one used by Pozdeeva
\textit{et al.} \cite{PozdeevaEtAl2016} in a similar context.}:
\be
U_{eff}'(\Phi) =  \frac{ 4 \xi \kappa^2 \Phi U(\Phi) +(1-\xi \kappa^2 \Phi^2) U'(\Phi)}{1-\xi \kappa^2  (1-6 \xi)\Phi^2}
\label{DEffPotGenXi}
\ee

\section{General Considerations for conformal coupling $\xi=\frac16$}
\setcounter{equation}{0}
We will further specialize in this work to a conformally coupled scalar i.e. $\xi = 1/6$ which obviously results considerable simplifications.
In that case the Ricci scalar may be expressed in terms of the scalar field through the potential only, while the field derivative term drops:
\be
R=\kappa^2 \left ( 4 U(\Phi) -\Phi U'(\Phi) \right)~.
\label{RicciPhi}
\ee
So the equation of the scalar field is written in terms of the effective potential
\be
\nabla_{\mu} \nabla^{\mu} \Phi + U_{\rm eff}'(\Phi) = 0~,
\label{FEqScalarEffPot}
\ee
where the effective potential is now defined in terms of its derivative by a much simpler relation:
\be
U_{\rm eff}'(\Phi) =(1-\frac{\kappa^2}{6} \Phi^2) U'(\Phi) +  \frac{2\kappa^2}{3}  \Phi U(\Phi)
\label{EffPot}
\ee
At this point we note that since it is $U_{\rm eff}'(\Phi)$ which appears in the equation of motion for the scalar field $\Phi$, we can think of the field $\Phi$ as if it were minimally
coupled to gravity with (effective) potential given by $U_{\rm eff}(\Phi)$ (instead of
$U(\Phi)$).

\begin{figure}[t]
\begin{center}
{\includegraphics[width=7.8cm]{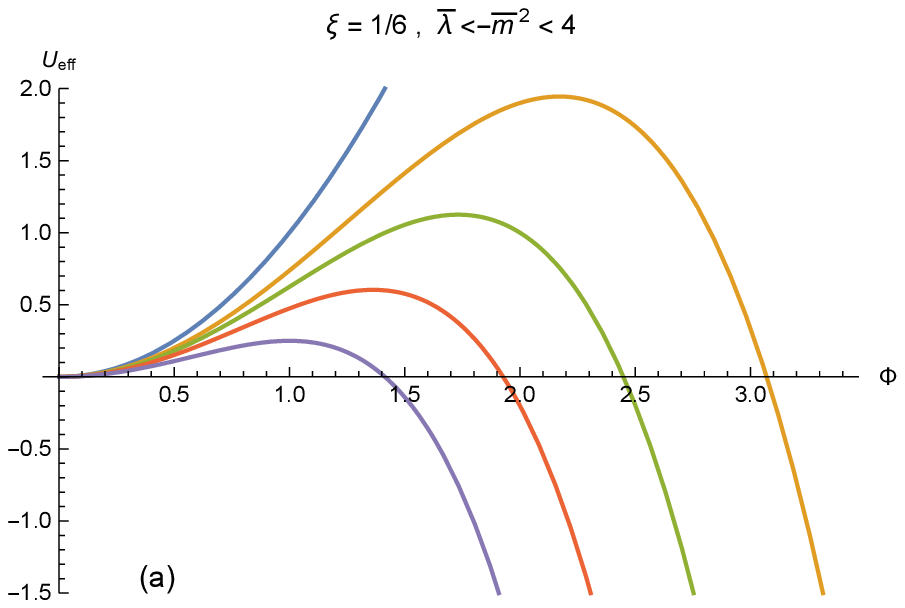}~~
\includegraphics[width=7.8cm]{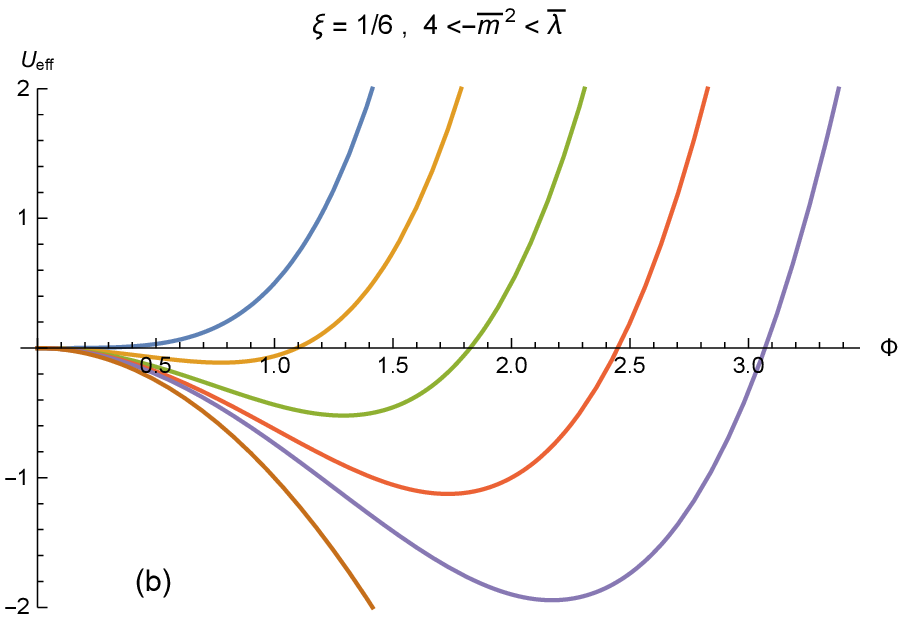}}
\caption{ \small{The rescaled effective potential $\bar{U}_{\rm eff}(\bar{\Phi})$ is shown
for $\xi=1/6$, $\nu=0$ and various parameters $\bar{m}^2$ .
(a) the parameter $\bar{\lambda}$ is fixed to $\bar{\lambda}=2$,
while the mass parameter $\bar{m}^2$ (see the definition in Eq.(\ref{DimensionlessCC}))
takes the values: $-\bar{m}^2=2,\, 2.35, \,2.5, \,2.7, \,3, \,4$.
In order to identify the corresponding curves, note that  $\bar{U}_{eff}(\bar{\Phi})$
decreases with $-\bar{m}^2$. These effective potentials can be obtained for potentials
\emph{bounded from below} due to the non-zero parameter $\bar{m}^2<0$ satisfying the inequalities
\eqref{CondUeffMaximumDmlss} or equivalently \eqref{condUeff}. Our bouncing solutions are
constructed with the effective potentials displayed on the left panel;
(b) $\bar{\lambda}=6$ and $-\bar{m}^2=4,\, 4.75, \,5.25, \,5.5, \,5.65, \,6$ so
these values violate the inequalities \eqref{CondUeffMaximumDmlss} or equivalently \eqref{condUeff}.
}}
\label{FigUeff}
\end{center}
\end{figure}

In the BGPS models, the effective potential has a shape similar to those shown in
Figure \ref{FigUeff}(a). Hence one may ask whether it is possible to start with a potential which
is bounded from below still yielding an effective potential which has the same shape.
%
%
For example, for an even polynomial of the sixth order we have
\bea
U(\Phi)&=&U_0+\frac{m^2}{2} \Phi^2  + \frac{\lambda}{4} \Phi^4 +
             \frac{\nu}{6} \Phi^6 \;\; , \label{Pot6th}\\
U_{\rm eff}(\Phi) &=& \left (\frac{m^2}{2}+ \frac{\kappa^2 U_0}{3}\right)\Phi^2  +
       \frac{1}{4}\left (\lambda + \frac{\kappa^2 m^2}{6} \right)\Phi^4 +
                  \frac{\nu}{6} \Phi^6 -\frac{\kappa^2\nu}{144} \Phi^8  \label{EffPot6th} .
\eea
The scalar field dynamics
is best understood from the shape of $U_{\rm eff}$, rather
than from the ``bare'' potential $U(\Phi)$ itself. Actually the potential in the Einstein frame (EF) is similar to the effective potential in the Jordan frame (JF) \cite{BoisseauEtAl2} so this is another way to understand the scalar field dynamics. We will return to this point
later.

In order that generic bounce solutions exist, it is the \textit{effective} potential which
should have a maximum at some non-zero $\Phi$, and may be unbounded from below as happens in
BGPS. The ``bare'' potential itself may be completely ``normal'' and still produce an effective
potential which will be unbounded from below as seen in Fig.\ref{FigUeff}(a) which corresponds
to the choice $\lambda>0$ and $\nu=0$ in Eqs.(\ref{Pot6th})--(\ref{EffPot6th}).


Restricting ourselves in the following to polynomial potentials of order four, therefore
setting $\nu=0$ in \eqref{EffPot6th}, inspection of \eqref{EffPot6th} gives immediately
an elegant answer: we must have a positive factor in front of $\Phi^2$, and a negative one
in front of $\Phi^4$. In other words, we get the condition
\be
\frac{m^2}{2} + \frac{\kappa^2 U_0}{3} > 0~,
\ee
and
\be
\lambda + \frac{\kappa^2 m^2}{6} < 0~.
\ee
These inequalities can be summarized as follows
\be  
-\frac{2\kappa^4 U_0}{3} <  \kappa^2 m^2 < -6\lambda~. \lb{condUeff}
\ee
The inequalities \eqref{condUeff} constitute the key result of this section, they allow
to generalize the bouncing solutions based on potentials unbounded from below, as in
BGPS, to quartic potentials which are \emph{bounded from below} when $\lambda>0$. It will be convenient
to reformulate these inequalities later using dimensionless variables, see
Eq.\eqref{CondUeffMaximumDmlss}.

The effective potentials shown on Fig.\ref{FigUeff}(a) (left panel) will be used to construct bouncing solutions.
They correspond to potentials $U$ bounded from below though these effective potentials have a shape similar to that
of the BGPS effective potential. The latter is obtained by substituting $m=0$ (and $\nu=0$), $U_0 >0$
and $\lambda <0$ in Eqs. (\ref{Pot6th})--(\ref{EffPot6th}).
Indeed, if $m^2=0$, this is the only way to satisfy the
conditions $\eqref{condUeff}$ and we are led to the inverted (unbounded from below) potential considered in BGPS.
From this wider perspective, the BGPS potential is just a simple special member in a large
family of potentials which may give rise to bouncing cosmologies.

This is the clearest way to see why the BGPS potential can be modified easily to become
bounded from below and more natural. The simplest way to turn the BGPS potential itself (with
$\lambda<0$) bounded from below is to add an increasing  higher power which will turn the
potential function $U(\Phi)$ bounded from below. If the turning point is for large enough
field values, the bouncing solutions will not change their general behavior, since the
effective potential $U_{eff}(\Phi)$ will be still unbounded. The two functions of
Eqs.(\ref{Pot6th})-(\ref{EffPot6th}) for $\lambda<0$ and $\nu>0$ (take for simplicity $m=0$)
are a simple example for this. We note incidentally that the potential of Eq. (\ref{Pot6th})
is in much use for $Q$-ball construction \cite{FriedbergEtAl1976,Coleman1985,Lynn1988}.

Viewed from a different point of view, if the potential $U(\Phi)$ contains a \textit{positive}
contribution with a power higher than 4, the effective potential will necessarily become
unbounded from below as demonstrated in Eq. (\ref{EffPot6th}) for a contribution of power 6.
We will thus assume from now on that our ``bare'' potential is always bounded from below,
i.e. in the polynomial potential of Eq. (\ref{Pot6th}) the parameter $\nu$ must be
non-negative, and if it vanishes, then $\lambda >0$. In what follows we will often keep $\nu$
in the equations, but we will not go into extensive study of its role which is straightforward
but cumbersome. In other words, in most of our concrete calculations and demonstrations we
will take $\nu = 0$.

It turns out to be much more transparent to analyze the situation in terms of rescaled
dimensionless variables.
%
We will just substitute $\xi=1/6$ in the general relation for arbitrary $\xi$
%
\be
\bar{\Phi}=\sqrt{\xi\kappa^2}\Phi \;\; , \;\;
\bar{x}^\mu =\sqrt{\xi\kappa^2 U_0}x^\mu  \;\; , \;\;
\bar{m}=\frac{m}{\sqrt{\xi\kappa^2 U_0}} \;\; , \;\;
\bar{\lambda}=\frac{\lambda}{\xi^2\kappa^4 U_0}\;\; , \;\;
\bar{\nu}=\frac{\nu}{\xi^3\kappa^6 U_0}~,
  \label{DimensionlessCC}
\ee
in terms of which the ``bare'' and the effective potential become (for $\xi=1/6$)
\bea \label{Bare+EffPot6thDmlss}
\bar{U}(\bar{\Phi}) &=& \frac{U(\bar{\Phi})}{U_0}= 1+\frac{\bar{m}^2}{2}\bar{\Phi}^2  +
\frac{\bar{\lambda}}{4}\bar{\Phi}^4 + \frac{\bar{\nu}}{6} \bar{\Phi}^6  \\
\bar{U}_{\rm eff}(\bar{\Phi}) &=& \frac{U_{\rm eff}(\bar{\Phi})}{U_0} =
\left (\frac{\bar{m}^2}{2}+ 2\right)\bar{\Phi}^2  +
\frac{1}{4}\left (\bar{\lambda} + \bar{m}^2 \right)\bar{\Phi}^4 +
\frac{\bar{\nu}}{6} \bar{\Phi}^6 -\frac{\bar{\nu}}{24} \bar{\Phi}^8 \nonumber
\eea
One may get the dimensionless version of all other equations for the conformal coupling by
the formal substitutions $\kappa=6$ and  $U_0 = 1$ and putting ``bars'' over all other
quantities. We will use alternatively dimensionful and dimensionless variables according to convenience.
%
%
%
%
\section{Bounce Solutions for $\xi= 1/6$}
\setcounter{equation}{0}
In the previous section we were mainly concerned with the scalar field dynamics which could
lead to cosmological bouncing solutions. We have to see how the larger class of effective
potentials found earlier will affect the space-time dynamics when bouncing solutions are
produced.
For this cosmological simple framework we assume a homogeneous time dependent scalar field
$\Phi (t)$ evolving in a standard flat FLRW Universe with metric
\be
ds^2=g_{\mu\nu}dx^\mu dx^\nu= dt^2- a^2 (t) \left(dr^2 + r^2 d \Omega_2^2\right)~.
\label{RWmetric}
\ee
Using the potential function of Eq. (\ref{Pot6th}), the scalar field equation becomes
\be
\ddot{\Phi}  +\frac{3 \dot{a}}{a } \dot{\Phi} +\left(m^2+\frac{2 \kappa^2 U_0}{3}\right)\Phi
+\left(\lambda +\frac{\kappa^2  m^2}{6}\right)\Phi  ^3 +\nu \Phi  ^5 -
\frac{\kappa^2 \nu}{18}   \Phi  ^7 =0
  \label{CosmologPhiEq}
\ee
and the two Einstein equations (for $G^1_1=G^2_2=G^3_3$ and for $G^0_0$) are given in explicit
form by
\bea\nonumber
\left(\frac{3}{\kappa^2 }-\frac{\Phi ^2}{2}\right)\left(\frac{2\ddot{a}}{a} +
   \frac{\dot{a}^2}{a^2}\right) + \frac{\dot{a} }{a} \Phi \dot{\Phi } +\frac{\dot{\Phi}^2}{2}
\hspace{5cm}\\ \label{Einst11}
  -3 U_0 + \left(\frac{2 \kappa^2  U_0}{3}-\frac{m^2}{2}\right)\Phi ^2 +
   \left( \frac{\lambda }{4}+\frac{\kappa^2  m^2}{6} \right)\Phi ^4
   +\frac{\nu }{2} \Phi ^6-\frac{\kappa^2  \nu}{18}   \Phi ^8 =0 \\
  \left(\frac{3}{\kappa^2 }-\frac{\Phi ^2}{2}\right) \frac{\dot{a}^2 }{a^2} -
\frac{\dot{a} }{a} \Phi \dot{\Phi } -\frac{\dot{\Phi }^2}{2} -U_0-\frac{m^2 }{2}\Phi ^2 -
   \frac{\lambda  }{4}\Phi ^4-\frac{\nu }{6} \Phi ^6=0 \label{Einst00}
\eea
Another useful relation is the explicit form of Eq. (\ref{RicciPhi}) (or just the sum
(\ref{Einst11})+(\ref{Einst00})),
\be
R=6\left(\frac{\ddot{a}}{a}+\frac{\dot{a}^2}{a^2}\right) =
                    \kappa^2\left(4 U_0+m^2 \Phi ^2 -\frac{\nu }{3} \Phi ^6\right)~,
  \label{RicciPhi2}
\ee
which may be used alternatively as a more convenient field equation. Actually, for the case
$m=\nu=0$ (while $\lambda$ may be non-zero) the scalar field decouples as the expression for
$R$, Eq.\eqref{RicciPhi2}, yields a constant and it is possible to completely integrate
the equations of motion and get already a bouncing
universe tending in the asymptotic future to General Relativity (GR) with a cosmological
constant \cite{BoisseauEtAl1,BoisseauEtAl2,BoisseauEtAl3}.
More accurately, in this particular case $R$ is a constant at all times, but this feature
of $R$ approaching a positive constant, or $H$ tending to a constant, is characteristic to
all the bouncing solutions we will find here.
So our aim is to find the conditions for the existence of solutions to the field equations
which pass through a bounce point which we take to occur at $t=0$ by exploiting the freedom to
redefine the origin of the time coordinate. The solutions we are after, are therefore
solutions that obey the ``initial conditions'' $H(0)=0$ and $\dot{H}(0)>0$. From this we
notice that the bounce solutions are determined by the value of the scalar field at $t=0$,
$\Phi(0)$, since the value of $\dot{\Phi}(0)$ is determined (up to a sign) by
the first order constraint (\ref{Einst00}) to obey
\be
\dot{\Phi}(0)^2 + U(\Phi(0)) = 0~. \lb{Phi0}
\ee
If we want to solve directly for $a(t)$, we may take an arbitrary value for $a(0)$ due to the
symmetry under rescaling of $a(t)$. We take naturally $a(0)=1$.

One may also solve directly for $H(t)=\dot{a}/a$ using the following equation which is a
combination of the field equations (\ref{CosmologPhiEq})-(\ref{Einst00}) (compare also Eq.\eqref{Fr2} in what follows):
\be
\left(\frac{6}{\kappa^2 }-\Phi ^2 \right)\dot{H}+ H \Phi \dot{\Phi } + 2\dot{\Phi }^2 -
                                \Phi\ddot{\Phi}=0 \label{EinstHubbleEq}
\ee
Equation \eqref{EinstHubbleEq} will be very useful for some of our derivations.
From the scalar field equation Eq. (\ref{CosmologPhiEq}), we easily see that the main
properties of the scalar field solutions are determined by the effective potential
$U_{\rm eff}(\Phi)$ defined in Eq. (\ref{EffPot6th}) whose derivative appears in
Eq. (\ref{CosmologPhiEq}). In analogy with Newtonian mechanics we note that the scalar field
equation is similar to that of a point particle moving under the potential $U_{\rm eff}(\Phi)$
with an additional damping term with the Hubble function in the role of the viscosity
coefficient which may also be negative.

In particular, by inspection of the effective potential $U_{\rm eff}(\Phi)$ we find the possible asymptotic
values of $\Phi$. If $\Phi$ tends to a constant there are only two possibilities: (i)
$\Phi(t)\rightarrow 0$, (ii) $\Phi(t)$ goes at late times to the second critical point where
$U'_{\rm eff}(\Phi_{\ast})=0$,
%
%
The third possibility is that  $\Phi(t)$ diverges in a finite time after the bounce but these
are not interesting solutions.
The regular cases correspond to the following late time behavior of the Hubble function:
(i) $H(t)\rightarrow H_0=\sqrt{\kappa^2 U_0 /3}$ , (ii) $H(t)\rightarrow H_{\ast}=
\sqrt{2\kappa^2 U(\Phi_{\ast})/(6-\kappa^2 \Phi^2_{\ast})}$ ,

By time reflection symmetry, the analogous
possibilities exist for $t<0$ as well as in the future if we change $t\rightarrow -t$;
we however break ``by hand'' the symmetry by the requirement that $\dot{H} (0)>0$ which
selects the realistic types of solutions with $H(t)>0$ for $t>0$.

In spite of the existence of the BGPS explicit (analytic) solution, it is unrealistic to
expect that analytic solutions can be found in general\footnote{See however
\cite{KamenshchikEtAl2015,KamenshchikEtAl2016} for some further analytical results in this
theory and \cite{PozdeevaVernov2017} for Induced Gravity.}. So the solutions of the field
equations which will be presented in this report were found numerically.
\begin{figure}[t]
\begin{center}
{\includegraphics[width=7.5cm]{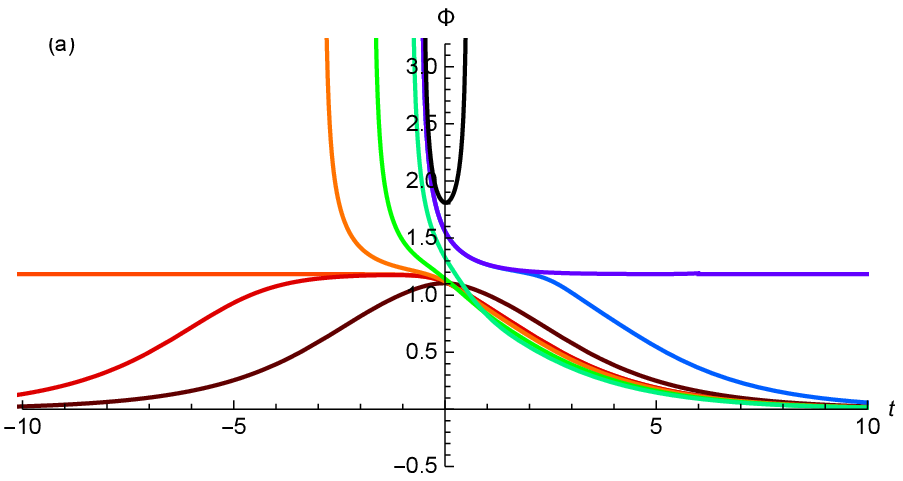}}
{\includegraphics[width=7.5cm]{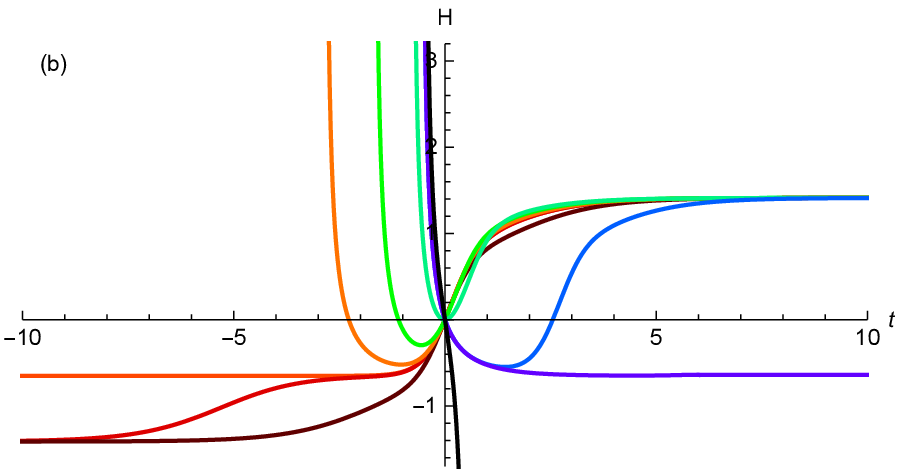}}
\caption{\label{ProfilesPhi+H} \small{Several representative ``time-profiles'' ($\bar{\Phi}(t)$
and $H(t)$) of cosmological bounce solutions for the conformally coupled case ($\xi=1/6$).
The parameter values are $\bar{m}^2=-2.25$, $\bar{\lambda}=1$ ($\nu=0$) which corresponds
to the upper (red) part of the allowed region of parameter space (see
Fig.\ref{FigParameterSpaceCC}). These models satisfy the inequalities \eqref{inequL3} and
\emph{all} bounce solutions violate the condition $F>0$ at the bounce, with
$\bar{\Phi}_-=1.104 > 1$.
The solutions displayed have the following $\bar{\Phi}(0)$ values: 1.104 = $\bar{\Phi}_-$, 1.114,
1.11437, 1.120, 1.145, 1.330, 1.560, 1.56025, 1.810.
Note that the symmetric $\bar{\Phi}$ solution has a maximum at the bounce. Note also a
symmetric $\bar{\Phi}$ solution with a minimum at $t=0$ with $\bar{\Phi}(0)>\bar{\Phi}_*$.
Bounce solutions are those that satisfy $H(0)=0$, $\dot{H}(0)>0$ and have no future
singularities.}}
\end{center}
\end{figure}

Solving the field equations
we find several kinds of solutions
which are shown in  Fig.\ref{ProfilesPhi+H} for the representative choice
$(\bar{\lambda}=1,\bar{m}^2=-2.25)$. They indeed fit the general framework described above.
Note that this figure demonstrates solutions which are not all bounce solutions. The
one with $\bar{\Phi}(0)=1.810$ has only a finite lifetime, starting from a Big Bang and ending
in a big crunch. The solution with $\bar{\Phi}(0)=1.56025$ is the limiting solution which
tends asymptotically to the maximum of the effective potential at
$\bar{\Phi}=\bar{\Phi}_{\ast}=1.18322$ (see Eq. (\ref{LocationUeffMaximum}) below). The
corresponding spacetime starts from a Big Bang, crosses a maximal expansion and contracts
exponentially. The third solution in Fig.\ref{ProfilesPhi+H} which does not satisfy the
initial conditions imposed above for a bounce, is actually a bounce solution with the only
``fault'' of having the bounce point not at $t=0$, but at some later time. At $t=0$ it is
still contracting. However, there is no need to include it as an independent bounce solution,
since it is related by time translation to one of the bounce solutions with smaller
$\bar{\Phi}(0)$ which also have $\dot{H}(0)>0$. So, the maximal value of $\bar{\Phi}(0)$ that
we should include in the $\bar{\Phi}(0)$ domain which produces bounce solution is the one
which corresponds to $H(0)=\dot{H}(0)=0$.

These solutions exist in different regions of parameter space, i.e. the $(\bar{\lambda} \, ,
\bar{m}^2)$ plane, which correspond to different relations between the maximum $\bar{\Phi}_*$
of $\bar{U}_{\rm eff}(\bar{\Phi})$ and the zeroes $\bar{\Phi}_-,~\bar{\Phi}_+$ of
$\bar{U}(\bar{\Phi})$. This analysis will be crucial for the assessment of possible bouncing
solutions.

The inequalities \eqref{condUeff} which are required in order to reproduce an effective
potential similar to that of the BGPS model, become in terms of our dimensionless
variables
%
\be
-4 <  \bar{m}^2 < -\bar{\lambda}~.
   \label{CondUeffMaximumDmlss}
\ee
The maximum itself is located at\footnote{This corresponds to two values because of the
symmetry of the potential. We will concentrate on $\Phi>0$.}
\be
\bar{\Phi}^2_{\ast}=-\frac{\bar{m}^2+ 4}{\bar{m}^2+\bar{\lambda}}~.
   \label{LocationUeffMaximum}
\ee
\begin{figure}[t!]
\begin{center}
{\includegraphics[width=7.9cm]{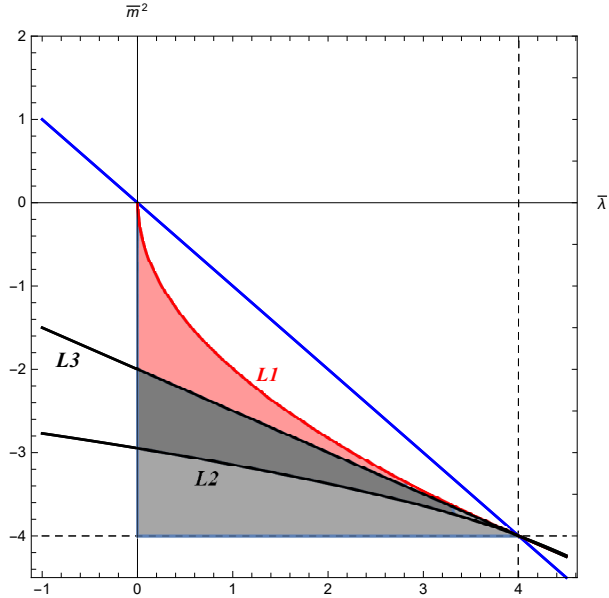}}
\caption{\small{The various regions of the $(\bar{\lambda}\, ,\bar{m}^2)$
plane where bounce solutions exist for a Higgs-like potential function $\bar{U}(\bar{\Phi})$,
i.e. a potential bounded from below, and an effective potential $\bar{U}_{\rm eff}(\bar{\Phi})$
which has a maximum (see Fig.\ref{FigUeff}). We take $\xi=1/6$.
The upper part of the allowed region between $L1$ and $L3$ (red) satisfies the inequalities
\eqref{inequL3}. Fig.\ref{ProfilesPhi+H} corresponds to this region. All bouncing
solutions in this region violate the condition $F>0$ at the bounce, but completely regular
($H$ and $\bar{\Phi}$ finite everywhere) bounce solutions do exist.
The lower part of the allowed region between $L2$ and $L3$ (black) satisfies the inequalities
\eqref{ineqlL3}. Fig.\ref{ProfilesPhi+DownHRegion} corresponds to this region. Bounce solutions exist in this region satisfying $F>0$ at the bounce till
the asymptotic future, but $H$ and $\bar{\Phi}$ become singular in the past for all these
solutions, regular bounce solutions do not exist.
The grey area at the bottom contains solutions with ``wrong'' asymptotic behavior.
For more details, see text.}}
\label{FigParameterSpaceCC}
\end{center}
\end{figure}
A second necessary condition is inferred from the first order constraint - Eq. (\ref{Einst00})
taken at $t=0$. The scalar field value $\Phi(0)$ at the bounce belongs to the domain where the
potential $U$ is negative
\be
U(\Phi(0))<0\;\;\;\;\ \Rightarrow\;\;\;\; \bar{\Phi}_{-}<\bar{\Phi}(0)<\bar{\Phi}_{+}~,
   \label{CondUnegative}
\ee
where
%
%
$\bar{\Phi}_{\pm}$ are the zeroes of $\bar{U}(\bar{\Phi})$:
\be
\bar{\Phi}^2_{\pm}=-\frac{\bar{m}^2}{\bar{\lambda}}
\pm\sqrt{\left(\frac{\bar{m}^2}{\bar{\lambda}}\right)^2-\frac{4}{\bar{\lambda}}}~.
   \label{Uzeroes}
\ee
We find therefore, that in addition to the condition (\ref{CondUeffMaximumDmlss}),
the parameters of the potential should also satisfy
\be
0<\bar{\lambda} < \bar{m}^4/4~.
   \label{ParSpaceSecondCond}
\ee
Finally, $F>0$ is guaranteed whenever $\bar{\Phi}<1$. Obviously, all field values
$\bar{\Phi}(0)$ at the bounce would satisfy $F>0$ for a given choice of parameters yielding
$\bar{\Phi}_+ < 1$.
However, in view of \eqref{Uzeroes}, \eqref{CondUeffMaximumDmlss}, we always have
\be
\bar{\Phi}_+ > 1~. \lb{Phi+}
\ee
Therefore, if $\bar{\Phi}_- < 1$, only part of the bouncing solutions
will satisfy $F>0$ namely those solutions obeying
\be
\bar{\Phi}_- \le \bar{\Phi}(0) < 1~.
\ee
Hence while it looks naively as if $\bar{\Phi}(0)$ is bound from above by $\bar{\Phi}_+$, the
upper zero of $\bar{U}$, it is actually bound by $1$ if we insist on $F>0$. So in this
respect the situation is similar to the BGPS solutions where the (inverted) potential $U$
has only one zero ($\bar{\Phi}_-$). Obviously, when $\bar{\Phi}_- > 1$, all existing bouncing
solutions necessarily violate the condition $F>0$.
So the condition $F>0$ brings additional important constraint on the model parameters.

Let us return to the $(\bar{\lambda} \, ,\bar{m}^2)$ parameter plane, see
Fig.\ref{FigParameterSpaceCC}.
If we insist on a potential bounded from below (i.e. $\bar{\lambda}>0$), the region in which
bounce solutions are found is the right triangular domain with the ``parabolic hypotenuse''
(which we will denote $L1$) shown in Fig.\ref{FigParameterSpaceCC}. The curve $L1$ expresses
the constraint \eqref{ParSpaceSecondCond}.
There is however an additional condition which originates from the asymptotic behavior of
the solutions which should correspond to a universe without future singularities. All the
solutions below the line $L2$ in Fig.\ref{FigParameterSpaceCC} have the ``wrong'' asymptotic
behavior, so this line is the lower edge of the domain of existence of bounce solutions: to
each point in this domain corresponds a family of bounce solutions characterized by the value
of $\Phi(0)$ which lies in the interval $(\Phi_{-} , \Phi_{+})$. Remember that by bounce
solutions we mean solutions which satisfy $H(0)=0$, $\dot{H}(0)>0$ and no future singularities.
The line $L2$ can be found only numerically and it is well approximated by
\be
 \bar{m}^2=-4+0.42(4-\bar{\lambda})^{2/3}
   \label{LineL2}
\ee
This domain of bounce solutions is further divided by the line $L3$ described by
\be
 \bar{m}^2=-2-\bar{\lambda}/2
   \label{ParSpaceCondsDmlssL3}
\ee
which separates the region where $\bar{\Phi}_{\ast}$ is located between  $\bar{\Phi}_{-}$ and
$\bar{\Phi}_{+}$ and the region where both $\bar{\Phi}_{-}$ and  $\bar{\Phi}_{+}$ are to
the right of $\bar{\Phi}_{\ast}$. These are the only two possibilities.
Actually $L3$ plays a crucial role.
Indeed, it is easily shown that points on the line $L3$ satisfy
\be
\bar{\Phi}_{-} = \bar{\Phi}_{*} = 1~. \lb{eqL3}
\ee
In the upper region between the line $L3$ and the curve $L1$, defined by
$\bar{m}^2>-2-\bar{\lambda}/2$,
one has the following inequalities
\be
1 < \bar{\Phi}_{-} < \bar{\Phi}_{*} < \bar{\Phi}_{+}~. \lb{inequL3}
\ee
Hence in this region \emph{all} quantities $\bar{\Phi}_{-}$, $\bar{\Phi}_{*}$, and not only
$\bar{\Phi}_{+}$, are larger than one.
Therefore, if we impose the condition $F>0$, this region of the parameter plane is excluded
as all bouncing solutions necessarily violate it (at least) at the bounce. This is a key result
of this section and we will return below to it.

\begin{figure}[t!]
\begin{center}
{\includegraphics[width=7.8cm]{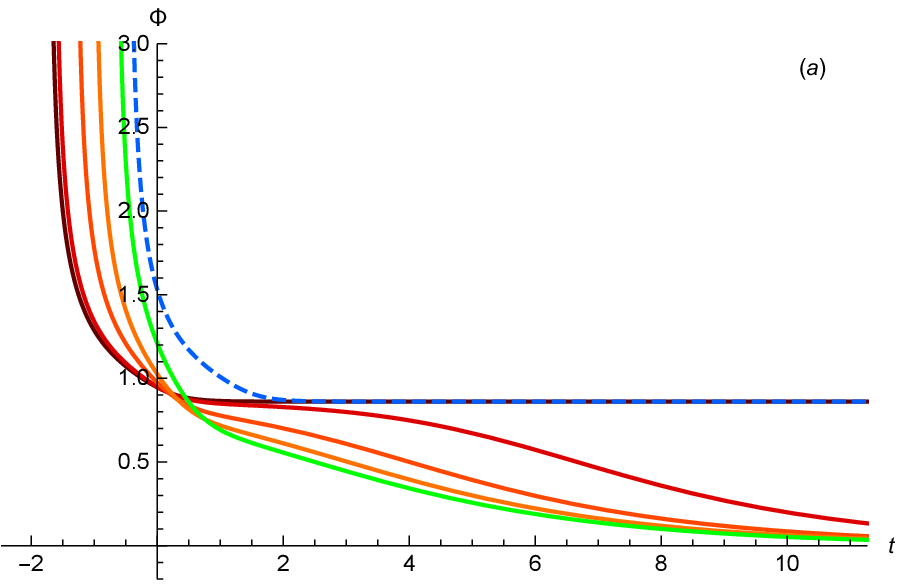}}
{\includegraphics[width=7.8cm]{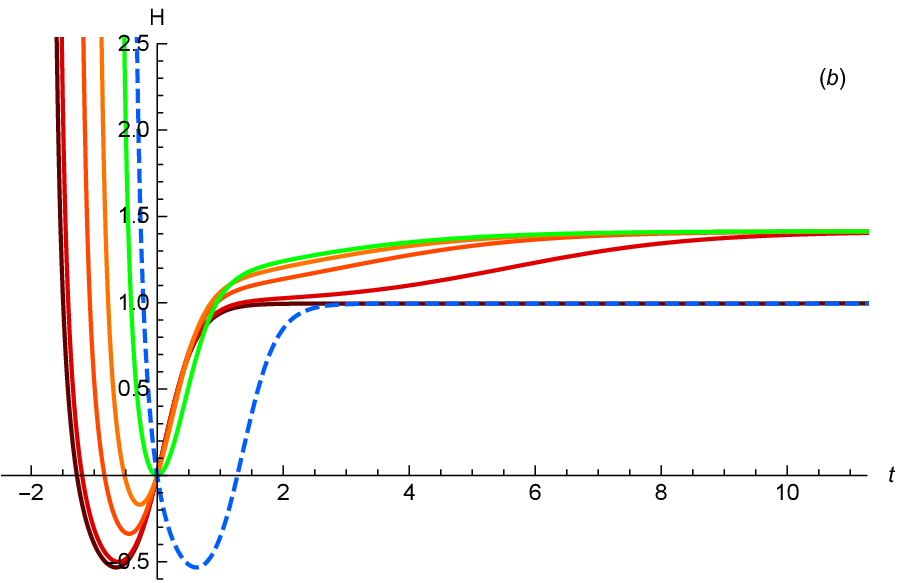}}
\caption{\small{Several representative ``time-profiles''
($\Phi(t)$ and $H(t)$) of cosmological solutions of the conformally coupled case ($\xi=1/6$)
in the lower part of the allowed region of parameter space (black), see
Fig.\ref{FigParameterSpaceCC}. These models satisfy the inequalities \eqref{ineqlL3}.
The parameters values are $\bar{m}^2=-2.7225$ and $\bar{\lambda}=1$ ($\nu=0$). The
$\bar{\Phi}(0)$ values are: 0.94672, 0.9500, 0.9750, 1.025, 1.210, 1.53142. Bounce solutions
do exist which do not violate the condition $F>0$ at the bounce. However \emph{all} bounce
solutions become singular in the past, regular bounce solutions cannot exist.
Note that the largest $\bar{\Phi}(0)$ solution (dashed curve) does not fulfill the
$\dot{H}(0)>0$ condition and is equivalent up to a time translation to the solution with
the smallest $\bar{\Phi}(0)$.}}
\label{ProfilesPhi+DownHRegion}
\end{center}
\end{figure}

In the lower region (between $L2$ and $L3$) defined by $\bar{m}^2 < -2-\bar{\lambda}/2$,
one has
\be
\bar{\Phi}_{*} < \bar{\Phi}_{-} < 1 < \bar{\Phi}_{+}~. \lb{ineqlL3}
\ee
So in this region of the parameter plane part of the existing bounce solutions do satisfy
the condition $F>0$ at (and after) the bounce. However, as we will show below $H$ cannot be
globally regular, this region allows only for partial bounce solutions with $H$ diverging,
and becoming positive, before the bounce.
We are now in a position to confirm analytically the curves displayed on Figures
\ref{ProfilesPhi+H}, \ref{ProfilesPhi+DownHRegion} and to exhibit the differences with the
BGPS solutions.
While the scalar field dynamics is best understood in terms of the effective potential
$U_{\rm eff}$ one must actually return to the potential $U$ itself in order to analyze the
space-time dynamics. To see this, it is convenient to consider the Friedmann equations
in a more compact form\footnote{Eq. \eqref{Fr1} is actually the time component of the Einstein equations of Eq. \eqref{FEqScalar+Grav_Gen} for the flat FLRW metric \eqref{RWmetric}, while Eq. \eqref{Fr2} is a combination of the time and space components. Compare also Eq. \eqref{EinstHubbleEq}.}
\bea
3 F H^2 &=& \frac12 \dot{\Phi}^2 - 3 H \dot{F} + U~, \label{Fr1}\\
-2 F \dot{H} &=& \dot{\Phi}^2 + \ddot{F} - H \dot{F}~. \label{Fr2}
\eea
and apply them to the present case with the choice \eqref{F} and \eqref{Z=1}.
Let us consider a bouncing solution where $\Phi$
is very large before the bounce so that $F$ becomes negative. In order to be able to pass the
bounce, the scalar field must obviously decrease in this region so $\dot{\phi}$ must be
negative. We can immediately find a striking difference between an inverted and non-inverted
(polynomial) potential considered in this work. Indeed, looking more closely at
\eqref{Fr1}, it is seen that while the last term is positive for a non-inverted potential,
it is clear that Eq.\eqref{Fr1} is satisfied provided we have
\be
- 3 H \dot{F} < 0~, \lb{condH}
\ee
which is our case forces $H$ to be \emph{positive}.
In other words, we cannot have a globally contracting universe before the bounce for a
non-inverted (polynomial) potential. This shows also the impossibility for a bounce
solution to satisfy globally $\dot{H}>0$.
In the BGPS model this was possible precisely because $U$ was negative for large $\Phi$. So
while the scalar field dynamics is essentially similar with the potentials considered here,
the space-time dynamics on the other hand can show crucial differences.
From this discussion, it follows that the only way one can hope to have a globally
growing Hubble function $H$, also before the bounce when the field $\Phi$ becomes
large, is to have a potential $U$ which becomes negative rather than positive for large $\Phi$.

Of course, there is still the option that the field $\Phi$ itself does not become large before
the bounce. Such a solution would be regular before as well as after the bounce. As we have
seen from the scalar field dynamics, this is possible if $\Phi$ tends either to zero
or to $\Phi_*$ before and after the bounce.

Using our previous analysis of the $(\bar{\lambda} \, ,\bar{m}^2)$ parameter plane, we note
that several situations can arise. In the full allowed region between $L1$ and $L2$, we get
from \eqref{Uzeroes} the inequality
\be
\bar{\Phi}_{-} > 0~. \lb{ineqPhimin}
\ee
Therefore a solution tending to zero in the past and in the future must
have a local maximum at some finite time in order to satisfy
$\bar{\Phi}(0) \ge \bar{\Phi}_- > 0$.
Using \eqref{F},\eqref{Fr2} or \eqref{EinstHubbleEq} in its explicit form, the following
equality is easily obtained at any extremum of $\Phi$
\be
6 F \dot{H} - \Phi \ddot{\Phi} = 0~. \lb{Fneg}
\ee
Assuming globally the inequality $\dot{H}>0$, and not just at the bounce, we see that a
maximum is possible only when $F<0$. Several solutions of this kind are displayed on Fig.\ref{ProfilesPhi+H}. There is still the possibility that the solution tends asymptotically
to $\bar{\Phi}_*$ in the past and/or in the future. This does not modify our conclusions.
Indeed, in view of the inequalities \eqref{inequL3},\eqref{ineqlL3}, either one has a local
maximum at some finite time when the parameters are in the lower part of the allowed region
between lines $L2$ and $L3$ (see Fig.\ref{FigParameterSpaceCC}); either a maximum can be
avoided in principle for $\bar{\Phi}_* > \bar{\Phi}_-$ which is possible only when the
parameters lie in the upper part of the allowed region between lines $L3$ and $L1$ and
both $\bar{\Phi}_*,~\bar{\Phi}_-$ violate $F>0$, see \eqref{inequL3}.

Note the existence of a symmetric solution with respect to time reversal satisfying
$\dot{\Phi}_0 = 0$ and $\ddot{\Phi}_0 <0$ at the bounce, see Fig.\ref{ProfilesPhi+H}.
The corresponding Hubble function has a ``step-like'' shape tending to a positive constant
as $t\rightarrow\infty$ and a negative constant as $t\rightarrow -\infty$. That is
exponential contraction until a minimal value of the scale
factor followed by an exponential expansion.
Such a realization implies in particular $\Phi(0)=\Phi_-$ from \eqref{Fr1}. Hence for such
a solution to exist, the model parameters must be chosen such that $\bar{\Phi}_- > 1$. In
other words, this is only possible in the upper part of the allowed region, between lines
$L1$ and $L3$. The parameter values of Fig.\ref{ProfilesPhi+H} correspond indeed to this
region. Such a symmetric solution cannot exist in the lower part between lines $L2$ and $L3$,
see e.g. the parameters chosen for Fig. \ref{ProfilesPhi+DownHRegion}.

In our derivation of the existence of regular bounce solution in the upper region between
$L3$ and $L1$ however violating the condition $F>0$, we have used the fact that
$\Phi_-$ is either larger or smaller than one together with the assumption
that the background evolution obeys $\dot{H}>0$.
Actually it is possible to generalize these
results. Indeed, from the dynamics of the scalar field $\bar{\Phi}$ and the shape of the
effective potential, it is clear that any extremum of $\bar{\Phi}$ corresponds either to a
maximum $\bar{\Phi}_{\rm max}$ when
\be
\bar{\Phi}_{\rm max} < \bar{\Phi}_*~, \lb{max}
\ee
or to a minimum $\bar{\Phi}_{\rm min}$ when
\be
\bar{\Phi}_{\rm min} > \bar{\Phi}_*~. \lb{min}
\ee
It is then immediately seen that the maximum of a regular bounce solution in the upper region
between $L3$ and $L1$ (satisfying \eqref{inequL3}), will necessarily violate the condition
$F>0$ as we have in this case
\be
1 < \bar{\Phi}_{-} \le \bar{\Phi}_{\rm max} < \bar{\Phi}_*~.
\ee
We stress that this result is now obtained without any assumption concerning the behaviour
of $H$. Of course at the maximum $\bar{\Phi}_{\rm max}$, we necessarily have $\dot{H}>0$ from
\eqref{Fneg}. Such solutions can and do exist.

Looking at the lower region between $L2$ and $L3$ (satisfying \eqref{ineqlL3}), it is seen
that regular bouncing solutions cannot exist at all. It is interesting that in this case
bounce solutions which do not violate the condition $F>0$ do exist, and some examples are
displayed on Fig.\ref{ProfilesPhi+DownHRegion}, but regular solutions do not exist.
One could summarize the situation in the following way: either we can avoid violation of the
condition $F>0$ but no regular solution exists, or we can construct regular bounce solutions
but we necessarily violate the condition $F>0$.

Fig.\ref{ProfilesPhi+DownHRegion} depicts the behavior of a representative family of solutions
corresponding to $(\bar{\lambda}=1,\bar{m}^2=-2.7225)$ (lower region between $L2$ and $L3$).
The asymptotic values are given as before by
Eqs(\ref{AsymptBehaviorProfilesI})--(\ref{AsymptBehaviorProfilesII})
which give in the particular case of Fig.\ref{ProfilesPhi+DownHRegion} the two possibilities:
($\bar{\Phi}(t) \rightarrow  0$ , $\bar{H} (t) \rightarrow \sqrt{2}$) and
($\bar{\Phi}(t) \rightarrow 0.86119$, $\bar{H} (t) \rightarrow 0.9952$).

For the sake of completeness we note that monotonically expanding solutions exist all over
the place: they are the only kind where bounces cannot exist (i.e. no $\dot{a}=0$) and coexist
with bounce solutions where they are possible. These expanding solutions start at a Big Bang
and become typically exponentially growing because of the $U_0>0$ parameter.
The ``time-profiles'' of Fig.\ref{ProfilesPhi+H} correspond to the point $(\bar{\lambda}=1,
\bar{m}^2=-2.25)$ which lies in the upper region between $L1$ and $L3$. The solutions satisfy
the appropriate asymptotic conditions of the above mentioned types: type (i):
\be
\bar{\Phi}(t)\rightarrow 0 \;\; , \;\;\;\;   \bar{H}^2 (t) \rightarrow \bar{H}^2_{0} = 2
\label{AsymptBehaviorProfilesI}
\ee
and type (ii):
\be
\bar{\Phi}(t)\rightarrow \bar{\Phi}_{\ast}=\sqrt{-\frac{\bar{m}^2+ 4}{\bar{m}^2 +
\bar{\lambda}}} \;\; , \;\;\;\;   \bar{H}^2 (t) \rightarrow \bar{H}^2_{\ast} =
\frac{4\bar{\lambda} - \bar{m}^4}{2(\bar{\lambda} +\bar{m}^2)}
\label{AsymptBehaviorProfilesII}.
\ee
For the particular case of Fig.\ref{ProfilesPhi+H} we have for type (ii) the numerical values:
$\bar{\Phi}(t) \rightarrow 1.18322$, $\bar{H} (t) \rightarrow -0.65192$.
Fig.\ref{ProfilesPhi+H} contains also one representative of type (iii) which ``lives'' for a
finite time between a Big Bang and a big crunch.

To summarize, it is possible for part of the parameter region to allow for globally regular
bouncing solutions, however such solutions necessarily violate the condition $F>0$.
While this is not problematic at the homogeneous background level, it was shown in
\cite{Starobinsky1981} that anisotropies would become infinite when $F\to 0$. Hence in such
a scenario, primordial fluctuations can be generated only after the bounce, and more precisely
after the quantity $F$ crosses zero (upwards) for the second time. In this respect, the
situation is quite similar to the standard inflationary picture in a Big Bang universe.

%
%
Finally we note the correspondence with the BGPS solutions.
The crucial difference is that since $m=0$, there is only one (positive) root to the
equation $U(\Phi)=0$, which is $\Phi^2_1 =\sqrt{-4U_0 /\lambda}$.
It is clear that we have $\lambda<0$, and hence also $\bar{\lambda}<0$, for $U_0 > 0$.
%
%
%
We must impose $\bar{\Phi}^2_1 < 1$ to ensure that the condition $F>0$ is not violated
at the bounce, or $\bar{\lambda}<-4$, which as expected from our results is only possible
for $\Phi^2_{\ast} < \Phi^2_1$ .
In that case regular bouncing solutions do not exist. The parameters were chosen in
\cite{BoisseauEtAl1} in this way so this corresponds to what happens in the lower region
between $L3$ and $L1$.

\section{Einstein frame}

\begin{figure}[t!]
\begin{center}
{\includegraphics[width=8cm]{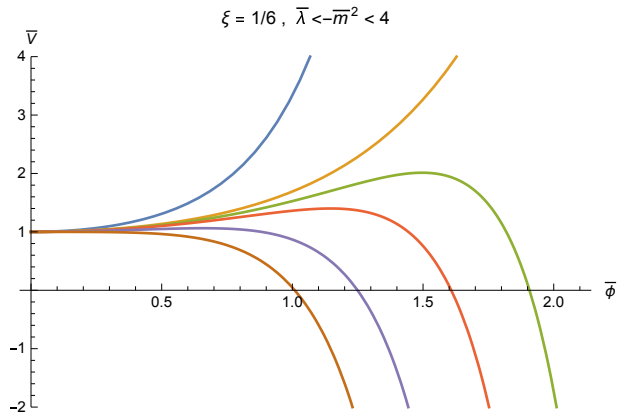}}
\caption{\small{The rescaled potential function in the Einstein frame $\bar{V}(\bar{\phi})$ is shown
for $\xi=1/6$, $\nu=0$ and $\bar{\lambda}=2$.
The mass parameter $\bar{m}^2$
takes the values: $-\bar{m}^2=2,\, 3, \,3.1, \,3.2, \,3.5, \,4$.
In order to identify the corresponding curves, note that  $\bar{V}(\bar{\phi})$
decreases as $-\bar{m}^2$ increases. From top to bottom: the first curve corresponds to the (red) upper
region in parameter space, $V$ never becomes negative and no bounce satisfying $F>0$ is possible for these parameters; the second curve is on L3; the other curves correspond to the lower (black) region for which $V$ can be negative allowing for a bounce satisfying $F>0$.}}
\label{V_EinsteinFrame}
\end{center}
\end{figure}

It is interesting to recover our results and to see how these are expressed in the Einstein
frame.
It is well-known that it is possible to transform the Lagrangian \eqref{actionJ} written in
the Jordan frame (JF), to a new one written in the Einstein frame (EF) as follows
\be
L= \frac{R_*}{2 \kappa ^2} + \frac12 g_*^{\mu\nu}\partial_{\mu}\phi\partial_{\nu}\phi-
V(\phi)~,
\ee
where the asterisk denotes the geometrical quantities in the EF; $\phi$ and $V$ are respectively the Einstein frame scalar field and potential. One can go from the
Jordan frame to the Einstein frame by means of the following transformations (for $Z=1$)
\bea
\left(\frac{d \phi}{d\Phi}\right)^2 &=& \frac{2}{\kappa ^2}\left(\frac{3}{4}
                    \left(\frac{d F/d \Phi}{F}\right)^2+\frac{1}{2F}\right)~,\nonumber\\
g^*_{\mu\nu} &=& \kappa^2 F(\Phi) ~g_{\mu\nu} = A^{-2}(\phi) ~g_{\mu\nu} \nonumber\\
V(\phi) &=& \kappa ^{-4}U(\Phi)F^{-2}(\Phi)  \lb{VVphi}
\eea
These transformations are obviously ill-defined when $F\le 0$.
We easily find for our system
\bea
\Phi &=&  \frac{\sqrt{6}}{\kappa}\tanh\left( \frac{ \kappa }{\sqrt{6}} \phi \right)\lb{Phitophi}\\
%
%
A(\phi) &=& \cosh\left(\frac{\kappa\phi}{\sqrt{6}}\right)~.\lb{A}
\eea
We have in particular that $\Phi\ge 0$ is a monotonically growing function of $\phi$ with
$\bar{\Phi}\to 1$ for $\phi\to \infty$.
Using \eqref{VVphi} we easily obtain the potential $V(\phi)$, viz.
\be
V(\phi) = U_0\cosh^4\left(\frac{\kappa}{\sqrt{6}}\phi \right) +
\frac{3m^2}{\kappa^2}\cosh^2\left(\frac{\kappa}{\sqrt{6}}\phi \right)
                                      \sinh^2\left(\frac{\kappa}{\sqrt{6}}\phi \right) +
\frac{9\lambda}{\kappa^4}\sinh^4\left(\frac{\kappa}{\sqrt{6}}\phi \right)~.\lb{VV}
\ee
This potential is even and its behaviour depends crucially on the parameter values.
We can now introduce dimensionless quantities $\bar{\phi},~\bar{V}$ in the same way
as we did earlier in the JF, see \eqref{DimensionlessCC}, and write \eqref{VV} in the
appealing form (see Fig. \ref{V_EinsteinFrame})
\be
\bar{V}(\bar{\phi}) = \cosh^4(\bar{\phi}) +
\frac{\bar{m}^2}{2}\cosh^2(\bar{\phi})\sinh^2(\bar{\phi}) +
\frac{\bar{\lambda}}{4}\sinh^4(\bar{\phi})~. \lb{bVV}
\ee
Remembering the different regions in parameter space, see Fig.\ref{FigParameterSpaceCC},
we see that $\bar{V}$ tends asymptotically to $+\infty$ in the (upper red) region between $L3$ and $L1$ and on the line $L3$,
while $V$ tends to $-\infty$ in the (lower black) region between $L2$ and $L3$.

The crucial fact is that only in the last case does the potential become negative.
It is straightforward to check from \eqref{VVphi} that when the EF is well-defined,
the EF potential $V$ is negative when the JF potential $U$ is negative.
As the existence of a bounce solution in the JF requires that $U$ becomes negative -- in our case
between $\Phi_-$ and $\Phi_+$ -- around the bounce, it is clear that only the (lower black) region between $L2$
and $L3$ can produce bounce solutions satisfying $F>0$ around the bounce.
The limit is the line $L3$ on which $\bar{\Phi}_* = \bar{\Phi}_- = 1$ or
$\bar{\phi}_* = \bar{\phi}_- = \infty$, and $V = \left( 1 - \frac{\lambda}{4} \right) \cosh^2(\bar{\phi}) + \frac{\lambda}{4} > 0$.
In the upper region including the line $L3$, bounce solutions satisfying $F>0$ cannot exist at all because
$V$ is non negative there.
This confirms the results found in section 3, namely that while bounce solutions do exist in the JF
when the parameters lie in the upper region between $L3$ and $L1$, these solutions violate the condition
$F>0$ around the bounce. Our analysis in the EF confirms that this violation is unavoidable, which is the key result
of this section.

We add now some more material which helps in understanding the correspondence between both frames.
We know that a spatially flat Robertson-Walker (RW) metric in the JF yieds a spatially flat RW
metric in the EF, viz.
\be
d s_*^2 = d t_*^2 - a_*^2(t_*)d\vec{ x}^2~,
\ee
where we have defined $dt =  A(\phi) dt_*$ and $a = A(\phi)a_*$. Note that
$t_*$ is a monotonically growing function of $t$ in the region $F>0$. We need of course the Hubble function in the EF which is obviously $H_*  = (da_*/dt_*)/a_*$.
So, the Friedmann equations in the EF are given by
\bea
\frac{3}{\kappa^2} H_*^2 &=& \frac{1}{2} \left(\frac{d\phi}{d t_*}\right) ^2  + V \lb{EFFr1}\\
\frac{1}{\kappa^2}\frac{d H_*}{d t_*} &=& -\frac{1}{2} \left(\frac{d\phi}{d t_*}\right)^2
\lb{EFFr2}
\eea
while the Klein-Gordon equation for the scalar field is
\be
\frac{d^2\phi}{dt_*^2} + 3 H_* \frac{d\phi}{dt_*} + \frac{d V}{d\phi} = 0~.
\ee
Note that from \eqref{EFFr2} we cannot get a bounce in the EF because $\frac{d H_*}{d t_*}\le 0$.
We can further easily derive the following equalities
\be
\frac{d\phi}{dt_*} = \cosh^3 \left( \frac{\kappa}{\sqrt{6}}\phi \right) ~\frac{d\Phi}{dt}~,\lb{dphitodPhi}
\ee
and
\be
H = A^{-1} \left[ H_* + \frac{ \kappa }{\sqrt{6}}
         \tanh\left( \frac{ \kappa }{\sqrt{6}} \phi \right) \frac{d\phi}{dt_*}\right]~. \lb{HtoH*}
\ee
Eq.\eqref{dphitodPhi} shows that an extremum of $\Phi$ in the JF corresponds to an extremum of $\phi$ in the
EF.
One can show through Eq.\eqref{Phitophi} that the maximum of $V(\phi)$ corresponds to the maximum of $U_{\rm eff}(\Phi)$.
To illustrate the correspondence between both frames in a specific case, we note that in particular when
$U(\Phi_-)=0$, a bouncing solution is possible with $\dot{\Phi}\big|_0 = 0$, $\Phi_0 = \Phi_-$ at the bounce.
It is easily checked from \eqref{VVphi},\eqref{dphitodPhi} and \eqref{HtoH*}, that this
corresponds in the EF to $V(\phi_-)=0$, $\frac{d\phi}{dt_*}=0$ and $H_*=0$.

\section{Conclusion} \label{Concl}
\setcounter{equation}{0}

Cosmological bouncing universes without spatial curvature require violation of the null energy condition. Effective
violation of this condition and the construction of bouncing solutions is possible in
scalar-tensor models considered in this work. We have presented bouncing solutions which are
extensions of the BGPS solutions found earlier. The latter were based on potentials unbounded
from below.
As a first important result, considering quartic potentials, we show that if the potential parameters
satisfy the inequalities \eqref{CondUeffMaximumDmlss} or equivalently \eqref{condUeff},
an arbitrarily large family of potentials \emph{which are bounded from below} is obtained which can produce
bouncing solutions.

Like in the BGPS solutions, the scalar field $\Phi(t)$ diverges generically in the past (and hence the condition
$F>0$ is violated in the past) in the new models considered in this work too.
However, unlike the BGPS solutions, we have shown the impossibility to have a regular Hubble function $H(t)$ from the
asymptotic past to the asymptotic future, $H(t)$ will necessarily diverge in the past before the bounce, see eqs. \eqref{Fr1} - \eqref{condH}. This is due precisely to the fact
that the potential $U$ is unbounded from below in the BGPS model while it is positive for large $\Phi$ for the models considered in this work.
Actually the only way to avoid this divergence of $H(t)$ is to avoid altogether the divergence of $\Phi(t)$.
In that case, regular bouncing solutions with a step-like Hubble function $H$ can be constructed in part of the parameter space.
We have shown however
that such solutions necessarily violate the condition $F>0$ at the bounce, see the discussion in section 3 and in particular
the crucial inequalities  \eqref{inequL3}, \eqref{ineqlL3}.
We have found the regions in parameter space, see Fig.\ref{FigParameterSpaceCC},
where any of these solutions exist. All these possibilities are illustrated in Fig.\ref{ProfilesPhi+H} and Fig.\ref{ProfilesPhi+DownHRegion}.
These results were confirmed by looking at the scalar field potential in the Einstein frame, see Eq. \eqref{VV}  or \eqref{bVV}.
As a bouncing solution requires $V<0$ (at the bounce), we have found indeed that $V<0$ is only possible in the (lower black) region
between L2 and L3 of the parameter space. This corresponds indeed to the region found in the Jordan frame where bounce solutions
with $F>0$ exist.

To summarize, either we have bouncing solutions which are regular in the future and which do not violate the condition
$F>0$ around the bounce, but the scalar field $\Phi$ \emph{and also the Hubble function $H$} will diverge before the bounce;
or else we can have bouncing solutions for which both $\Phi$ and $H$ are globally regular, but these will violate the condition
$F>0$ around the bounce.
If such a scenario is realized in our Universe, it implies that the globally regular bouncing universes
should be perfectly isotropic around the bounce, in particular there can be no perturbations before the bounce. The introduction of positive spatial curvature can provide a way to avoid the problem of $F>0$ violation around the bounce, that we leave for future investigation.

\textbf{Acknowledgement}

\noindent
AAS was supported by the RSF grant 21-12-00130. YV acknowledges support by the OUI Research Authority.



\begin{thebibliography}{99}

\bibitem{Planck:2018vyg}
N.~Aghanim {\it et al.},
Astron.\ Astrophys. {\bf 641}, A6 (2020), Astron.\ Astrophys. {\bf 652}, C4 (2021)  (erratum).

\bibitem{BICEP:2021xfz}
P.~A.~R.~Ade {\it et al.},
Phys.\ Rev.\ Lett. {\bf 127}, 151301 (2021).


\bibitem{Starobinsky:1982mr}
A.~A.~Starobinsky, JETP Lett. {\bf 37}, 66 (1983).

\bibitem{Muller:1989rp}
V.~M\"uller, H.~J.~Schmidt and A.~A.~Starobinsky,
Class.\ Quant.\ Grav. {\bf 7}, 1163 (1990).

\bibitem{Muller:2017nxg}
D.~M\"uller, A.~Ricciardone, A.~A.~Starobinsky and A.~V.~Toporensky,
Eur.\ Phys.\ J.\ C {\bf 78}, 311 (2018).

\bibitem{Mishra:2019ymr}
S.~S.~Mishra, D. M\"uller and A.~V.~Toporensky, Phys.\ Rev.\ D {\bf 102},
063523 (2020).

\bibitem{S78}
A.~A.~Starobinsky, Sov. Astron. Lett. {\bf 4}, 82 (1978).

\bibitem{Page:1984qt}
Don~N.~Page, Class.\ Quant.\ Grav. {\bf 1} 417 (1984).

\bibitem{Kamenshchik:1997szb}
A.~Yu.~Kamenshchik, I.~M.~Khalatnikov and A.~V.~Toporensky, Int.\ J.\ Mod.\ Phys.\ D {\bf 6}, 673 (1997).

\bibitem{Kamenshchik:1998xx}
A.~Yu.~Kamenshchik, I.~M.~Khalatnikov and A~V.~Toporensky, Int.\ J.\ Mod.\ Phys.\ D {\bf 7}, 129 (1998).

\bibitem{Kamenshchik:1998ix}
A.~Yu.~Kamenshchik, I.~M.~Khalatnikov, S.~V.~Savchenko and A.~V.~Toporensky, Phys.\ Rev.\ D {\bf 59}, 123516 (1999).

 \bibitem{Mukhanov-Brandenberger1991}
  V.~F.~Mukhanov and R.~H.~Brandenberger,
  Phys.\ Rev.\ Lett.\  {\bf 68}, 1969 (1992).

\bibitem{Novello-Bergliaffa2008}
  M.~Novello and S.~E.~P.~Bergliaffa,
  Phys.\ Rept.\  {\bf 463}, 127 (2008).

  \bibitem{BrandenbergerPeter2016}
  R.~Brandenberger and P.~Peter,
  Found.\ Phys.\  {\bf 47}, 797 (2017).

\bibitem{Barrow:2017yqt}
J.~D.~Barrow and C.~Ganguly, Phys.\ Rev.\ D {\bf 95}, 083515 (2017).

\bibitem{Bramberger:2019zez}
S.~F.~Bramberger and J.-L.~Lehners, Phys.\ Rev.\ D {\bf 99}, 123523 (2019).

\bibitem{Ganguly:2019llh}
C.~Ganguly and M.~Bruni, Phys.\ Rev.\ Lett. {\bf 123}, 201301 (2019).

\bibitem{Gungor:2020fce}
\"O.~G\"ung\"or and G.~D.~Starkman, JCAP {\bf 2104}, 003 (2021).

\bibitem{Boisseau:2000pr}
B.~Boisseau, G.~Esposito-Farese, D.~Polarski and A.~A.~Starobinsky, Phys.\ Rev.\ Lett. {\bf 85}, 2236 (2000).

   \bibitem{NojiriOdintsov2006}
  S.~Nojiri and S.~D.~Odintsov,
  Int.\ J.\ Geom.\ Meth.\ Mod.\ Phys.\  {\bf 4}, 115 (2007).

   \bibitem{CapozzielloDeLaurentis2011}
  S.~Capozziello and M.~De Laurentis,
  Phys.\ Rept.\  {\bf 509}, 167 (2011).

\bibitem{NojiriEtAl2017}
  S.~Nojiri, S.~D.~Odintsov and V.~K.~Oikonomou,
  Phys.\ Rept.\  {\bf 692}, 1 (2017).

  \bibitem{Veneziano1991}
  G.~Veneziano,
  Phys.\ Lett.\ B {\bf 265}, 287 (1991).

  \bibitem{Khoury+3-2001}
  J.~Khoury, B.~A.~Ovrut, P.~J.~Steinhardt and N.~Turok,
  Phys.\ Rev.\ D {\bf 64}, 123522 (2001).

  \bibitem{Khoury+4-2001}
  J.~Khoury, B.~A.~Ovrut, N.~Seiberg, P.~J.~Steinhardt and N.~Turok,
  Phys.\ Rev.\ D {\bf 65}, 086007 (2002).

 \bibitem{BoisseauEtAl1}
  B.~Boisseau, H.~Giacomini, D.~Polarski and A.~A.~Starobinsky,
  JCAP {\bf 1507}, 002 (2015).

\bibitem{BoisseauEtAl2}
  B.~Boisseau, H.~Giacomini and D.~Polarski,
  JCAP {\bf 1510}, 033 (2015).

  \bibitem{BoisseauEtAl3}
  B.~Boisseau, H.~Giacomini and D.~Polarski,
  JCAP {\bf 1605}, 048 (2016).

    \bibitem{KamenshchikEtAl2015}
  A.~Y.~Kamenshchik, E.~O.~Pozdeeva, A.~Tronconi, G.~Venturi and S.~Y.~Vernov,
  Class.\ Quant.\ Grav.\  {\bf 33}, 015004 (2016).

  \bibitem{PozdeevaEtAl2016}
  E.~O.~Pozdeeva, M.~A.~Skugoreva, A.~V.~Toporensky and S.~Y.~Vernov,
  JCAP {\bf 1612}, 006 (2016).

   \bibitem{Starobinsky1981}
  A.~A.~Starobinsky,
  Sov. Astron. Lett. {\bf 7}, 36 (1981).

  \bibitem{ABGS03} L. R. Abramo, L. Brenig, E. Gunzig, A. Saa,
  Phys. Rev.D {\bf 67}, 027301 (2003)

  \bibitem{Futamase-Maeda1987}
  T.~Futamase and K.~I.~Maeda,
  Phys.\ Rev.\ D {\bf 39}, 399 (1989).

  \bibitem{FutamaseEtAl1989}
  T.~Futamase, T.~Rothman and R.~Matzner,
  Phys.\ Rev.\ D {\bf 39}, 405 (1989).

  \bibitem{CaputaEtAl2013}
  P.~Caputa, S.~S.~Haque, J.~Olson and B.~Underwood,
  Class.\ Quant.\ Grav.\  {\bf 30}, 195013 (2013).


 \bibitem{KamenshchikEtAl2016}
  A.~Y.~Kamenshchik, E.~O.~Pozdeeva, S.~Y.~Vernov, A.~Tronconi and G.~Venturi,
  Phys.\ Rev.\ D {\bf 94}, 063510 (2016).

\bibitem{PozdeevaVernov2017}
  E.~O.~Pozdeeva and S.~Y.~Vernov,
  Phys.\ Part.\ Nucl.\  {\bf 49}, 914 (2018).

  \bibitem{FriedbergEtAl1976}
  R.~Friedberg, T.~D.~Lee and A.~Sirlin,
  Phys.\ Rev.\ D {\bf 13}, 2739 (1976).

\bibitem{Coleman1985}
  S.~R.~Coleman,
  Nucl.\ Phys.\ B {\bf 262}, 263 (1985);
  Erratum: [Nucl.\ Phys.\ B {\bf 269}, 744 (1986)].

  \bibitem{Lynn1988}
  B.~W.~Lynn,
  Nucl.\ Phys.\ B {\bf 321}, 465 (1989).

\end{thebibliography}
\end{document}